\documentclass{appolb}
\usepackage{epsfig}

\newcommand{\bDelta}{\mbox{\boldmath $\Delta$}}

\newcommand{\bkappa}{\mbox{\boldmath $\kappa$}}
\newcommand{\bp}{\mbox{\boldmath $p$}}

\newcommand{\bk}{\mbox{\boldmath $k$}}
\newcommand{\bb}{\mbox{\boldmath $b$}}

\newcommand{\bM}{\mbox{\boldmath $M$}}

\def\lsim{\mathrel{\rlap{\lower4pt\hbox{\hskip1pt$\sim$}}
    \raise1pt\hbox{$<$}}}         
\def\gsim{\mathrel{\rlap{\lower4pt\hbox{\hskip1pt$\sim$}}
    \raise1pt\hbox{$>$}}}         

\begin{document}
\title{Central exclusive production at high energies.%
\thanks{Presented at the Epihany 2009 conference:
"Hadron interactions at the dawn of the LHC", 5-7 January 2009; dedicated to the memory of Jan Kwieci\'nski.}%
}
\author{Antoni Szczurek
\address{Institute of Nuclear Physics PAN, PL-31-342 Cracow,
Poland,\\
University of Rzesz\'ow, PL-35-959 Rzesz\'ow, Poland
}
}
\maketitle
\begin{abstract}
I briefly review several mechanisms of central exclusive 
production of mesons at high energies. Some illustrative 
examples for the BNL RHIC, FNAL Tevatron and CERN LHC 
as well as for lower energies are discussed.
Some differential distributions are shown.
\end{abstract}

\section{Introduction}

\begin{figure}
\begin{center}
  \includegraphics[height=.25\textheight]{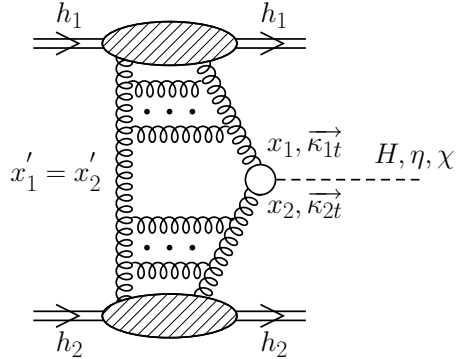}
\end{center}
  \caption{A sketch of the bare QCD mechanism of
exclusive heavy scalar meson production.
\label{fig:scalar_diagram}
}
\end{figure}

\begin{figure}
\begin{center}
  \includegraphics[height=.25\textheight]{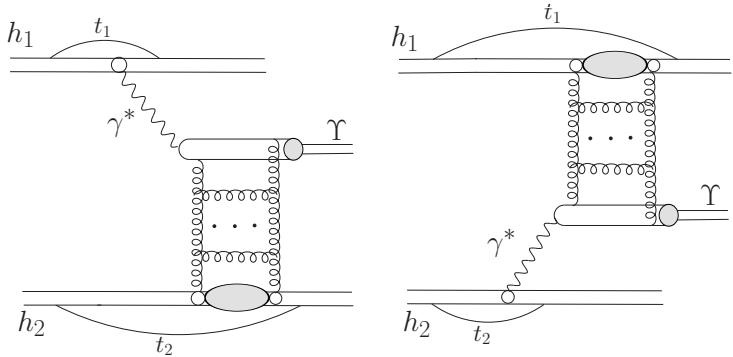}
\end{center}
  \caption{Two basic QED $\otimes$ QCD mechanisms of
exclusive heavy vector meson production.
\label{fig:vector_diagram}
}
\end{figure}

\begin{figure}
\begin{center}
  \includegraphics[height=.25\textheight]{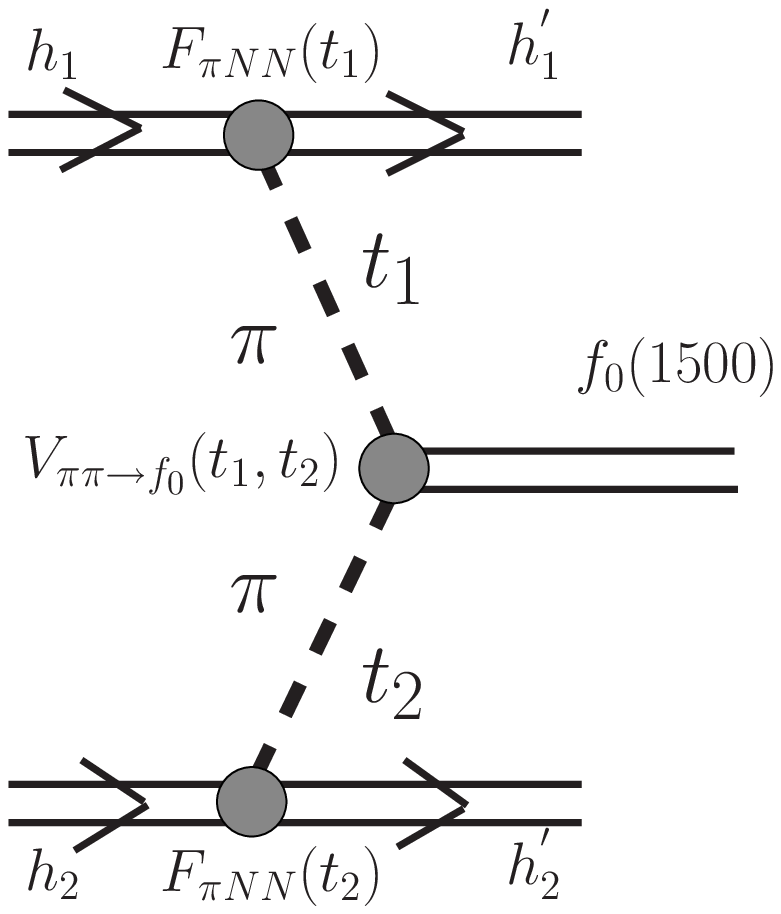}
\end{center}
  \caption{A sketch of the bare QCD mechanism of
exclusive heavy scalar $f_0(1500)$ meson production.
\label{fig:pion_pion_diagram}
}
\end{figure}

The exclusive production of mesons was studied in detail
mostly close to the kinematical threshold.
The Tevatron is a first accelerator which opens
a possibility to study the central (semi)exclusive 
production of mesons at high energies. 
A similar program will be carried out in the future 
at the LHC.
Here I review several mechanisms of exclusive
meson production studied recently by our group
(the details can be found in 
\cite{SPT07,SS07,PST08,PST09,RSS08,SL08,LS09,KSS09}).
In general, the mechanism of the reaction depends
on the quantum numbers of the meson and/or its internal
structure.
For heavy scalar mesons 
(scalar quarkonia, scalar glueballs) the mechanism of 
the production, shown in Fig.\ref{fig:scalar_diagram}, 
is exactly the same as for the diffractive Higgs boson 
production extensively discussed in recent years
by the Durham group \cite{KMR}.
The dominant mechanism for the exclusive heavy 
vector meson production is quite different.
Here there are two dominant processes shown in 
Fig.\ref{fig:vector_diagram}. When going to lower
energies the mechanism of the meson production becoming
more complicated and usually there exist more mechanisms. 
For example in Fig.\ref{fig:pion_pion_diagram} 
I show a new mechanism of the glueball production 
proposed recently in Ref.\cite{SL08}.

Other exclusive or semi-exclusive processes were
discussed during the conference by A. Martin \cite{Martin}, 
Ch. Royon \cite{Royon} and W. Guryn \cite{Guryn}.

\section{Selected examples}

Recently we have calculated differential cross sections 
for several exclusive processes:
\begin{itemize}
\item $ p p \to p p \eta' $, $ p p \to p p \eta_c$
 ( IP IP + $\gamma \gamma$ ) 
\item $ p p \to p p \chi_c(0^+)$ (IP IP + $\gamma \gamma$ ) 
\item $ p p \to p p \chi_c(1^+)$ (IP IP) 
\item $ p p \to p p f_0(1500)$ (IP IP + $\pi^+ \pi^-$)
\item $ p p \to p p J/\psi$ (IP $\gamma$ + $\gamma$ IP)
\item $ p p \to p p \Upsilon$ ( IP $\gamma$ + $\gamma$ IP)
\item $ p p \to p p \pi^+ \pi^-$ ( (IP + IR) $\otimes$ (IP + IR))
\item $ A A \to A A \rho^0 \rho^0 $ ($\gamma \gamma$)
\end{itemize}
Above the dominant mechanisms are shown in the parantheses.

The details of the formalism as well as a detailed analysis
of differential distribution in longitudinal and transverse
momenta can be found in our original papers
\cite{SPT07,SS07,PST08,RSS08,SL08,KSS09}).
Here I wish to discuss only some illustrative examples.

\subsection{Exclusive $\chi_c$ production 
in proton-proton and proton-antiproton collisions}

According to the Khoze-Martin-Ryskin approach (KMR) \cite{KMR}, we write
the amplitude of the exclusive double diffractive color singlet
production $pp\to pp\chi_{cJ}$ as
\begin{eqnarray}
{\cal
M}^{g^*g^*}=\frac{s}{2}\cdot\pi^2\frac12\frac{\delta_{c_1c_2}}{N_c^2-1}\,
\Im\int
d^2 q_{0,t}V^{c_1c_2}_J \nonumber \\
\frac{f^{off}_{g,1}(x_1,x_1',q_{0,t}^2,
q_{1,t}^2,t_1)f^{off}_{g,2}(x_2,x_2',q_{0,t}^2,q_{2,t}^2,t_2)}
{q_{0,t}^2\,q_{1,t}^2\, q_{2,t}^2} \; .
\label{ampl}
\end{eqnarray}
The amplitude is averaged over the color indices and
over the two transverse polarizations of the incoming gluons \cite{KMR}.

In calculating the vertex $V^{c_1c_2}_J$ we have included
off-shellness of gluons \cite{PST08}. The unintegrated
gluon distributions were taken from the literature.
We have demonstrated in Ref.\cite{PST08} that 
for relatively light $\chi_c(0)$, unlike for
the Higgs boson \cite{KMR}, the dominant 
contributions come from the nonperturbative regions 
of rather small gluon transverse momenta.

In Ref.\cite{PST08} we have made a detailed presentation
of differential distributions. Here only selected
results will be shown.
As an example I show distribution in Feynman variable 
$x_F$ of the $\chi_c$ meson for three different energies: 
W = 200 GeV (RHIC), W = 1960 GeV (Tevatron) and 
W = 14000 GeV (LHC) for different UGDFs from 
the literature.
Characteristic for central diffractive production all distributions
peak at $x_F \approx$ 0.
Although all UGDFs give a similar quality description of
the low-$x$ HERA data for the $F_2$ structure function,
they give quite different longitudinal momentum distributions of
$\chi_c(0^+)$.
The UGDFs which take into account saturation effects (GBW, KL) give much
lower cross section than the BFKL UGDF (dash-dotted line).
Therefore the process considered here would help, at least 
in principle, to constrain rather poorly known UGDFs.

\begin{figure}[!h] 
\includegraphics[width=4cm]{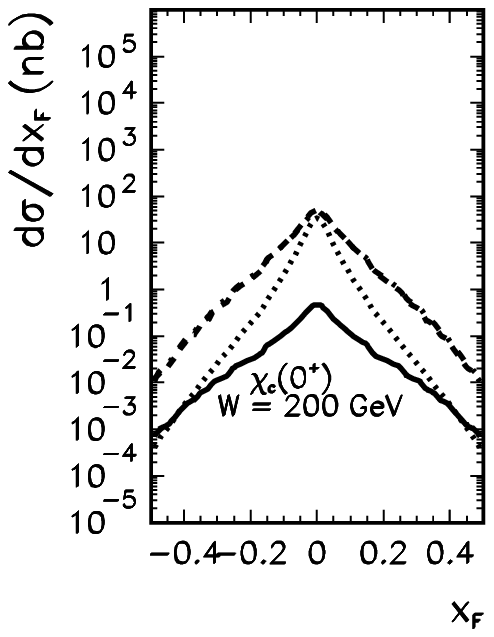}
\includegraphics[width=4cm]{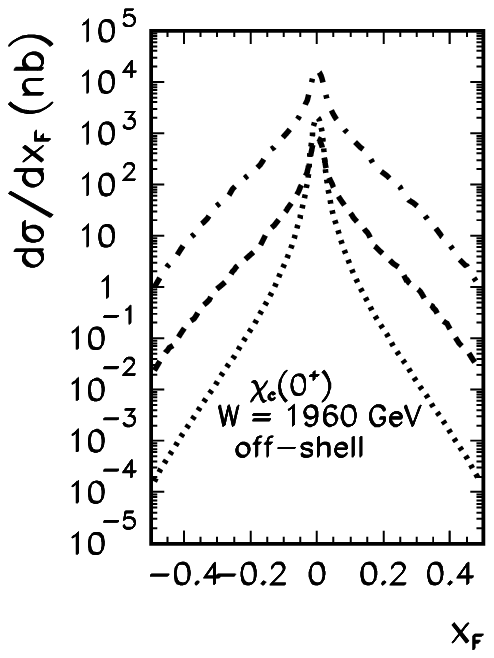}
\includegraphics[width=4cm]{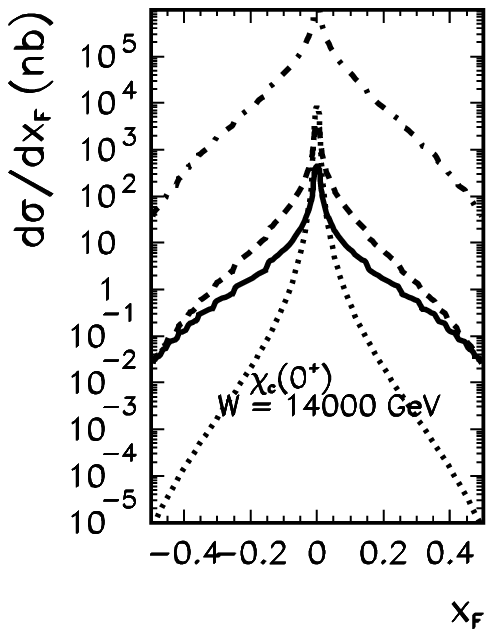}
   \caption{\label{fig:diffraction_ugdf}
Distribution in Feynman $x_F$ for RHIC (W=200 GeV), Tevatron (W=1960 GeV)
and LHC (W=14000 GeV) for different UGDFs: 
BFKL (dash-dotted), KL (dashed), GBW (dotted) and 
KMR (solid).}
\end{figure}

\begin{figure}[!h] 
\includegraphics[width=5cm]{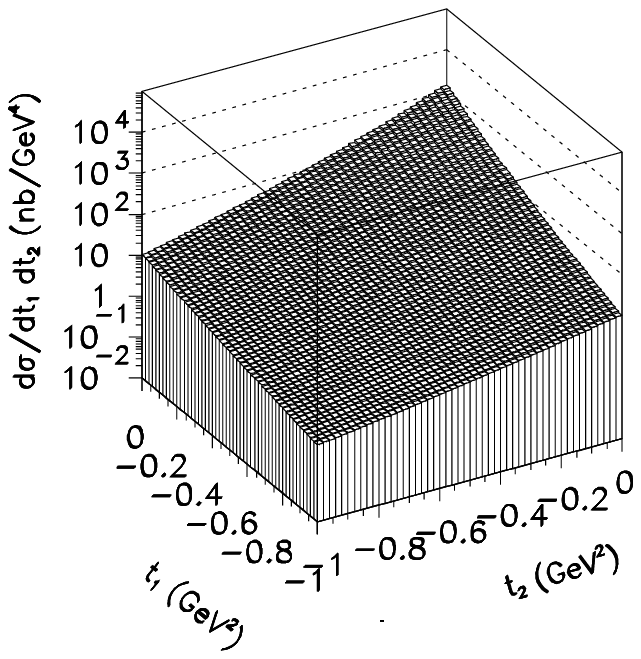}
\includegraphics[width=5cm]{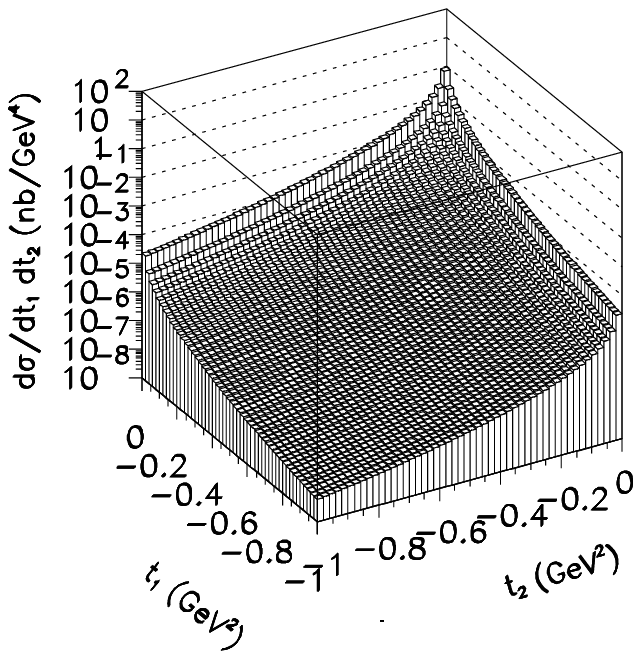}
   \caption{\label{fig:maps_t1t2}
Two-dimensional maps in $t_1$ and $t_2$ for
KL UGDF (left) and for two-photon fusion (right).
}
\end{figure}

The three-body reactions lead to correlations of outgoing
protons.
In Fig.\ref{fig:maps_t1t2} I show an example for 
diffractive mechanism with KL UGDF (as an example) 
as well as for photon-photon fusion \cite{PST08}.
The ($t_1,t_2$) distribution obtained in the photon-photon fusion
mechanism differs qualitatively from the distribution of 
the diffractive mechanism.
One can see a strong enhancement of the cross section when
$t_1$, $t_2$ $\to$ 0 which is caused by the photon propagators.
Although the diffractive
component is subjected to much stronger absorption effects than
the electromagnetic one, it is clear that the diffractive component 
dominates.

The situation with the axial-vector production is new compared to
both zero-spin case (scalar \cite{PST08}, pseudoscalar \cite{SPT07}
mesons) as well as to the vector meson production where the vector
meson is dominantly transversely polarized \cite{SS07,RSS08}, at
least for small transferred four-momenta in the nucleonic line. The
axial-vector meson can be polarized both
transversely and longitudinally. 

There is interesting theoretical aspect of the double diffractive
production of the $\chi_c(1^{+})$ meson. The coupling 
$g g \chi_c(1^{++})$ vanishes for on-shell
gluons (so-called Landau-Yang theorem). According to the original
Landau-Yang theorem \cite{LY_theorem} the symmetries under space
rotation and inversion forbid the decay of the spin-1 particle into
two (on-shell) spin-1 particles (two photons, two gluons). The same
is true for the fusion of two on-shell gluons. The symmetry
arguments cannot be strictly applied for off-shell gluons. 

In Ref.\cite{PST09} we have confirmed 
explicitly that the Landau-Yang theorem is violated by
virtual effects in diffractive production of $\chi_c(1^+)$ leading
to very important observational consequences. In our approach the
off-shell effects are treated explicitly. For comparison, in the
standard KMR approach the corresponding cross section would vanish
due to their on-shell approximation. The measurement of the cross
section can be therefore a good test of the off-shell effects and
consequently UGDFs used in the calculation.

\begin{figure}[!h]    
\includegraphics[width=0.4\textwidth]{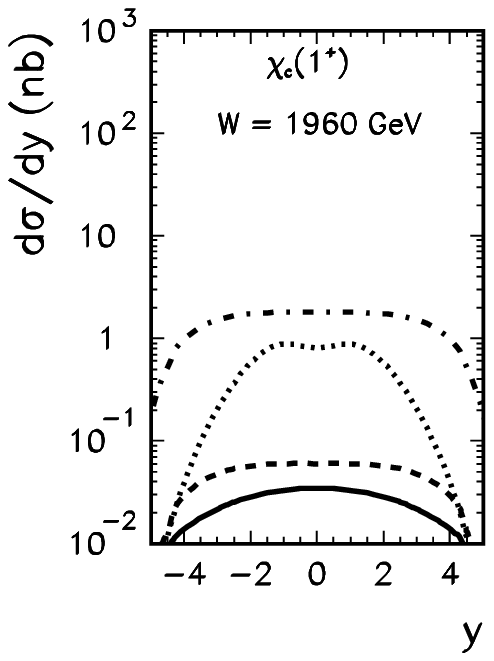}
\includegraphics[width=0.4\textwidth]{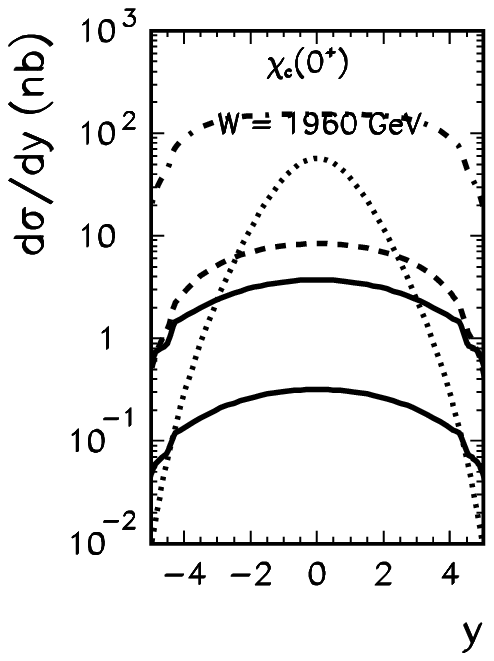}
   \caption{\label{fig:dsig_dy}
   \small Distribution in rapidity of $\chi_c(1^+)$ meson
(left panel) and $\chi_c(0^+)$ meson (right panel) for
different UGDFs.
}
\end{figure}

In Fig.~\ref{fig:dsig_dy} I show distributions
in rapidity $y$ for different UGDFs from the literature.
The results for different UGDFs differ significantly. The biggest
cross section is obtained with BFKL UGDF and the smallest cross
section with GBW UGDF. The big spread of the results is due to quite
different distributions of UGDFs in gluon transverse momenta
($q_{1t}, q_{2t}$), although when integrated over transverse momenta
distributions in longitudinal momentum fraction ($x_1, x_2$) are
fairly similar.

Comparing the left and right panels, the cross section for
the axial-vector $\chi_c(1^+)$ production is much smaller
(about two orders of magnitude)
than the cross section for the scalar $\chi_c(0^+)$
production. This is related to the
Landau-Yang theorem, which "causes" vanishing of
the cross section for on-shell gluons.
For axial-vector quarkonia the effect is purely
of off-shell nature and is due to the interplay of
the off-shell matrix element and off-diagonal UGDFs.
This interplay causes a huge sensitivity to UGDFs
observed in Fig.~\ref{fig:dsig_dy}.

At the Tevatron the $\chi_c$ mesons are measured through the $\gamma
+ J/\Psi$ decay channel. The axial-vector $\chi_c(1^+)$ meson has a
large branching fraction for radiative decay $\chi_c(1^+) \to \gamma
+ J/\psi$ (BR = 0.36 \cite{PDG}). This is much bigger than for the
scalar $\chi_c(0^+)$ where it is only about 1 \% \cite{PDG}.
Therefore the discussed off-shell efects are very important to
understand the situation in the $\gamma + J/\Psi$ channel observed
experimentally.

\subsection{Exclusive $f_0(1500)$ production \\
in proton-proton and proton-antiproton collisions}

In Ref.\cite{SL08} we have discussed exclusive
production of scalar $f_0(1500)$ in the following 
reactions:
\begin{eqnarray}
&&p + p \to p + f_0(1500) + p \; , \\
&&p + \bar p \to p + f_0(1500) + \bar p \; , \\
&&p + \bar p \to n + f_0(1500) + \bar n \; . 
\label{f0(1500)_reactions}
\end{eqnarray}
While the first process could be measured at the J-PARC 
complex being completed recently, 
the latter two reactions could be measured by 
the PANDA Collaboration
at the new complex FAIR planned in GSI Darmstadt.
The combination of these processes could shed more light
on the mechanism of $f_0(1500)$ production as well as 
on its nature.


\begin{figure}    %
\includegraphics[width=5cm]{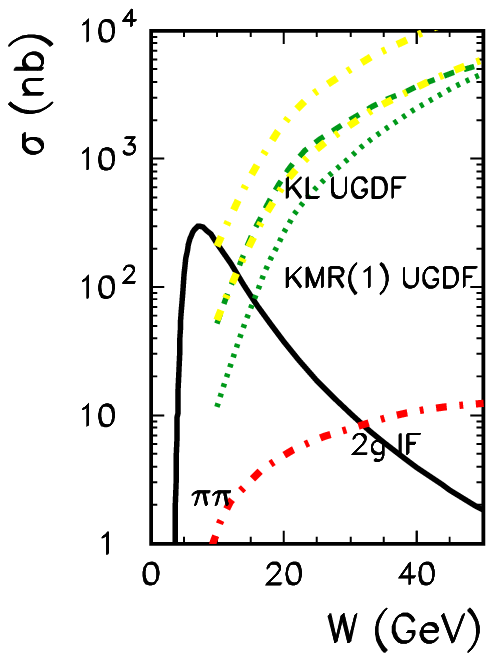}
\includegraphics[width=5cm]{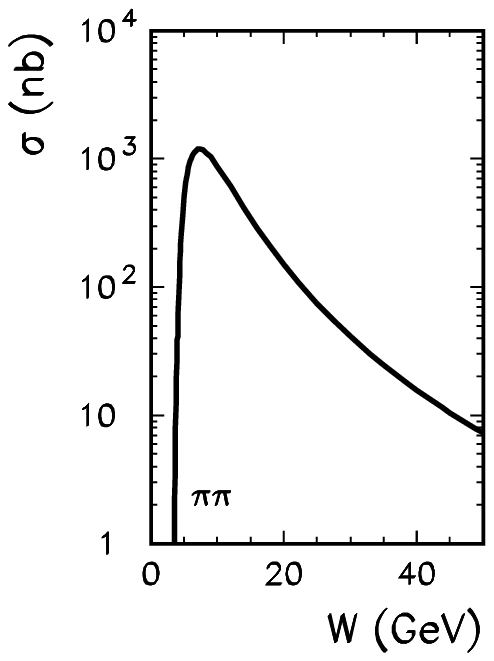}
\caption{\label{fig:sigma_W}
The integrated cross section as a function
of the center of mass energy for $p \bar p \to p \bar p f_0(1500)$ 
(left panel)
and $p \bar p \to n \bar n f_0(1500)$ (right panel) reactions.
The thick solid lines are for pion-pion MEC contribution 
($\Lambda$ = 0.8, 1.2 GeV), the dashed line is 
for QCD diffractive contribution obtained with the Kharzeev-Levin UGDF, 
the dotted line for the KMR approach
and the thin solid lines (blue on-line) are for "mixed" UGDF
(KL $\otimes$ Gaussian) with $\sigma_0$ = 0.5, 1 GeV. 
The dash-dotted line represents the two-gluon impact factor 
result. 
}
\end{figure}


In Ref.\cite{SL08} we have proposed a new mechanism 
(see Fig.\ref{fig:pion_pion_diagram}) which becomes 
dominant at lower energies.
In Fig.\ref{fig:sigma_W} we show the integrated cross section
for the exclusive $f_0(1500)$ elastic production
$p \bar p \to p f_0(1500) \bar p$ 
and for double charge exchange reaction
$p \bar p \to n f_0(1500) \bar n$.
The thick solid line represents the pion-pion component calculated with 
monopole vertex form factors with 
$\Lambda$ = 0.8 GeV (lower) and $\Lambda$ = 1.2 GeV (upper).
The difference between the lower and upper curves represents uncertainties
on the pion-pion component.
The pion-pion contribution grows quickly from the threshold, takes
maximum at $W \approx$ 6-7 GeV and then slowly drops with increasing
energy. The gluonic contribution calculated with unintegrated
gluon distributions drops with decreasing energy 
towards the kinematical threshold and seems to be about order of 
magnitude smaller than the pion-pion component at W = 10 GeV.
We show the result with Kharzeev-Levin UGDF (dashed line) which 
includes gluon saturation effects relevant for small-x, 
Khoze-Martin-Ryskin UGDF (dotted line) used for the exclusive 
production of the Higgs boson and the result with the 
"mixed prescription" (KL $\otimes$ Gaussian) \cite{SL08} 
for different values of the $\sigma_0$ parameter: 
0.5 GeV (upper thin solid line),
1.0 GeV (lower thin solid line). 
In the latter case results rather strongly depend on 
the value of the smearing parameter.

\subsection{Exclusive production of $\Upsilon$
\\
in proton-proton and proton-antiproton collisions}

\begin{figure}[!h]    %
\begin{center} 
\includegraphics[width=0.4\textwidth]{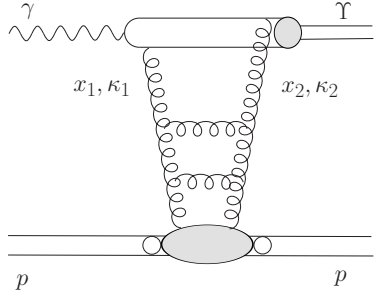}
\end{center}
   \caption{\label{fig:diagram_photon_pomeron}
   \small  A sketch of the exclusive $\gamma p \to \Upsilon p$ amplitude.}
\end{figure}

The photoproduction amplitude is the major
building block for our prediction of exclusive $\Upsilon$ production in hadronic collisions.
The amplitude for the reaction under consideration
is shown schematically in Fig.\ref{fig:diagram_photon_pomeron}.
As it is explained in Ref.\cite{INS06}, 
the imaginary part of the amplitude for the $\gamma^* p \to \Upsilon p$
process can be written as
\begin{eqnarray}
\Im m \; {\cal M}_{\lambda_{\gamma},\lambda_V}(W,t=-\bDelta^2,Q^2) =
W^2 \frac{c_\Upsilon \sqrt{4 \pi \alpha_{em}}}{4 \pi^2}
\int \frac{d^2\bkappa}{\kappa^4} \alpha_S(q^2)  
{\cal F}(x_1,x_2,\bkappa_1,\bkappa_2) 
\nonumber \\
\times
\int \frac{dz d^2 \bk}{z (1-z)}  
I_{\lambda_{\gamma}, \lambda_V}(z,\bk,\bkappa_1,\bkappa_2,Q^2) \; ,
\label{full_imaginary}
\end{eqnarray}
where the transverse momenta of gluons coupled to the $Q \bar Q$
pair can be written as
\begin{eqnarray}
\bkappa_1 = \bkappa + {\bDelta \over 2} \, , \, 
\bkappa_2 = - \bkappa + {\bDelta \over 2} \, .
\end{eqnarray}
The quantity ${\cal F}(x_1,x_2,\bkappa_1,\bkappa_2)$ is
the off diagonal unintegrated gluon distribution. 
Explicit expressions for $I_{\lambda_{\gamma}, \lambda_V}$ 
can be found in \cite{INS06}.
For heavy vector mesons, helicity--flip transitions may be neglected,
and we concentrate on the $s$--channel helicity conserving amplitude,
$\lambda_\gamma = \lambda_V$. In the forward scattering limit, i.e.
for $\bDelta =0$, azimuthal integrations can be performed analytically,
and we obtain the following representation for the
imaginary part of the amplitude for forward photoproduction $\gamma p \to \Upsilon p$ :
\begin{eqnarray}
\Im m \, {\cal M}(W,\Delta^2 = 0,Q^2=0) =
W^2 \frac{c_\Upsilon \sqrt{4 \pi \alpha_{em}}}{4 \pi^2} \, 2 \, 
 \int_0^1 \frac{dz}{z(1-z)}
\int_0^\infty \pi dk^2 \psi_V(z,k^2) \\
\int_0^\infty
 {\pi d\kappa^2 \over \kappa^4} \alpha_S(q^2) {\cal{F}}(x_{eff},\kappa^2)
\Big( A_0(z,k^2) \; W_0(k^2,\kappa^2) 
     + A_1(z,k^2) \; W_1(k^2,\kappa^2)
\Big) \, ,
\label{amplitude_forward}
\end{eqnarray}
where
\begin{eqnarray}
A_0(z,k^2) &=& m_b^2 + \frac{k^2 m_b}{M + 2 m_b}  \, , \\
A_1(z,k^2) &=& \Big[ z^2 + (1-z)^2 
    - (2z-1)^2 \frac{m_b}{M + 2 m_b} \Big] \, \frac{k^2}{k^2+m_b^2} \, ,
\end{eqnarray}
and
\begin{eqnarray}
W_0(k^2,\kappa^2) &=& 
{1 \over k^2 + m_b^2} - {1 \over \sqrt{(k^2-m_b^2-\kappa^2)^2 + 4 m_b^2 k^2}}
\, , 
\nonumber \\
W_1(k^2,\kappa^2) &=& 1 - { k^2 + m_b^2 \over 2 k^2}
\Big( 1 + {k^2 - m_b^2 - \kappa^2 \over 
\sqrt{(k^2 - m_b^2 - \kappa^2)^2 + 4 m_b^2 k^2 }}
\Big) \, .
\end{eqnarray}
We treat the $\Upsilon, \Upsilon'$ mesons as $b\bar{b}$ $s$--wave states,
the relevant formalism of light--cone wave functions 
is reviewed in \cite{INS06}.

\begin{figure}[!h]    %
\begin{center}
\includegraphics[width=0.8\textwidth]{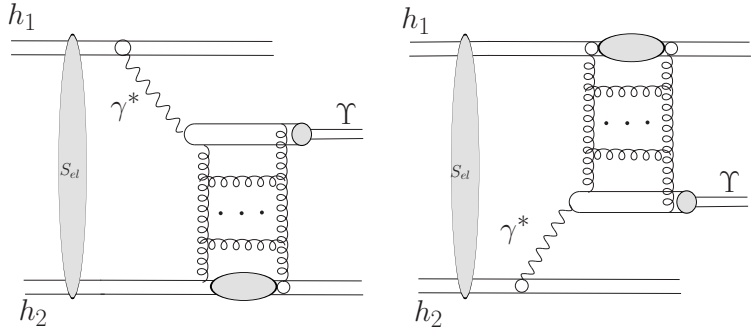}
\end{center}
   \caption{\label{fig:diagram_2}
   \small  A sketch of the two mechanisms considered:
photon-pomeron (left) and pomeron-photon (right), including absorptive corrections.}
\end{figure}

The necessary formalism for the calculation
of amplitudes and cross--sections was outlined in 
detail in Ref. \cite{SS07}. Here I give only a 
brief summary. 
The basic mechanisms are shown in Fig.\ref{fig:diagram_2}. 
The major difference from photoproduction at 
$ep$ machines, where the photon was emitted by a lepton
which does not participate in the strong interactions, now, both initial
state hadrons can be the source of the photon.  Therefore, it is now necessary
 to take account of the interference between two amplitudes. 
The photon exchange parts of the amplitude, involve only very small,
predominantly transverse momentum transfers. 
Here we concentrate on the kinematic domain, where 
the outgoing protons lose only tiny fractions 
$z_1,z_2 \ll 1$ of their longitudinal momenta, 
in practice $z \lsim 0.1$ means $y \lsim 3$.
In terms of the transverse momenta of outgoing hadrons,
$\bp_{1,2}$, the relevant four--momentum transfers are
$t_i = - (\bp_i^2 + z_i^2 m_p^2)/(1-z_i) \, , i = 1,2$,
and $s_1 \approx (1 -z_2) s$ and $s_2 \approx (1-z_1) s$ are the
familiar Mandelstam variables for the appropriate subsystems. 
Photon virtualities $Q_i^2$ are small (what counts here is that
$Q_i^2 \ll M_\Upsilon^2$), so that the contribution from 
longitudinal photons can be safely
neglected. Also, as mentioned above, we assume 
the $s$--channel--helicity conservation in the 
$\gamma^* \to \Upsilon$ transition.

The $2 \to 3$ Born-amplitude (without absorptive corrections) can be written in the form of a two--dimensional 
vector (corresponding to the two transverse (linear)
polarizations of the final state vector meson):
\begin{eqnarray} 
\bM^{(0)}(\bp_1,\bp_2) &&= e_1 {2 \over z_1} {\bp_1 \over t_1} 
{\cal{F}}_{\lambda_1' \lambda_1}(\bp_1,t_1)
{\cal {M}}_{\gamma^* h_2 \to V h_2}(s_2,t_2,Q_1^2)   
+ ( 1 \leftrightarrow 2 )
\end{eqnarray}
Inclusion of absorptive corrections (the 'elastic rescattering')
leads in momentum space to the full, absorbed amplitude
\begin{eqnarray}
\bM(\bp_1,\bp_2) &&= \int{d^2 \bk \over (2 \pi)^2} \, S_{el}(\bk) \,
\bM^{(0)}(\bp_1 - \bk, \bp_2 + \bk)  
= \bM^{(0)}(\bp_1,\bp_2) - \delta \bM(\bp_1,\bp_2) \, .
\nonumber \\
\label{rescattering term}
\end{eqnarray}
With 
\begin{equation}
S_{el}(\bk) = (2 \pi)^2 \delta^{(2)}(\bk) - \half T(\bk) \, \, \, ,
\, \, \, T(\bk) = \sigma^{p \bar p}_{tot}(s) \, \exp\Big(-\half B_{el} \bk^2 \Big) \, ,
\end{equation}
where $\sigma^{p \bar p}_{tot}(s) = 76$ mb, $B_{el} = 17 $ GeV$^{-2}$
were taken,
the absorptive correction $\delta \bM$ reads 
\begin{eqnarray}
\delta \bM(\bp_1,\bp_2) = \int {d^2\bk \over 2 (2\pi)^2} \, T(\bk) \,
\bM^{(0)}(\bp_1-\bk,\bp_2+\bk) \, .
\label{absorptive_corr}
\end{eqnarray}
The differential cross section is given in terms of $\bM$ as
\begin{equation}
d \sigma = { 1 \over 512 \pi^4 s^2 } | \bM |^2 \, dy dt_1 dt_2
d\phi \, ,
\end{equation}
where $y$ is the rapidity of the 
vector meson, and $\phi$ is the angle between $\bp_1$ and $\bp_2$.

\begin{figure}[!h]  
\includegraphics[width=0.45\textwidth]{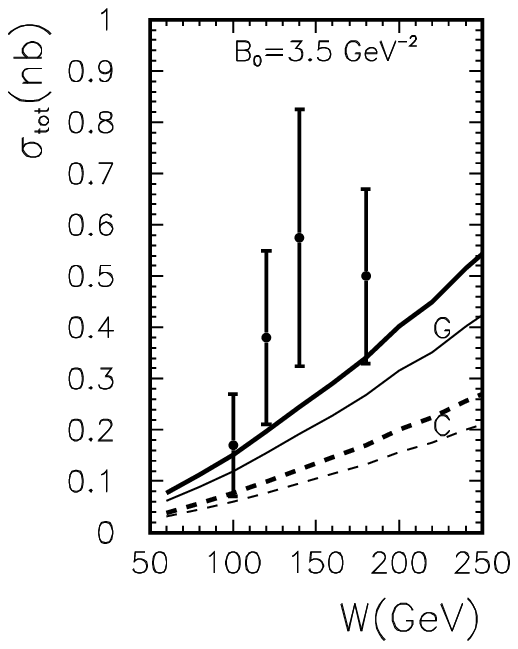}
\includegraphics[width=0.45\textwidth]{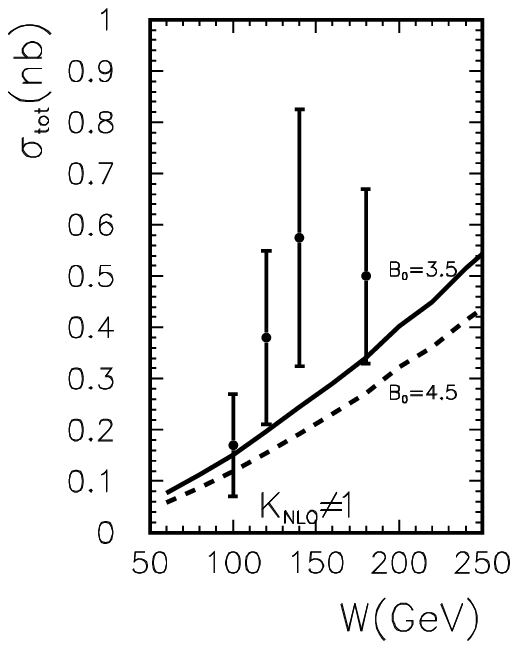}
   \caption{\label{fig:sig_tot}
   \small $\sigma_{tot} (\gamma p \to \Upsilon(1S) p)$ as a function
   of the $\gamma p$ cm--energy versus HERA--data. Left: dependence 
on the treatment of the $b \bar b \to \Upsilon$ transition; solid curves:
Gaussian (G) wave function, dashed curves: Coulomb--like (C) wave function.
Thick lines were obtained including the NLO--correction for the $\Upsilon$ decay width, while
for the thin lines $K_{NLO}=1$. Right: dependence on the 
slope parameter $B_0$ (given in GeV$^{-2})$, for the Gaussian wave function.
The experimental data are taken from \cite{ZEUS_old,H1,ZEUS_Upsilon}
}
\end{figure}
\begin{figure}[!h]  
\includegraphics[width=0.45\textwidth]{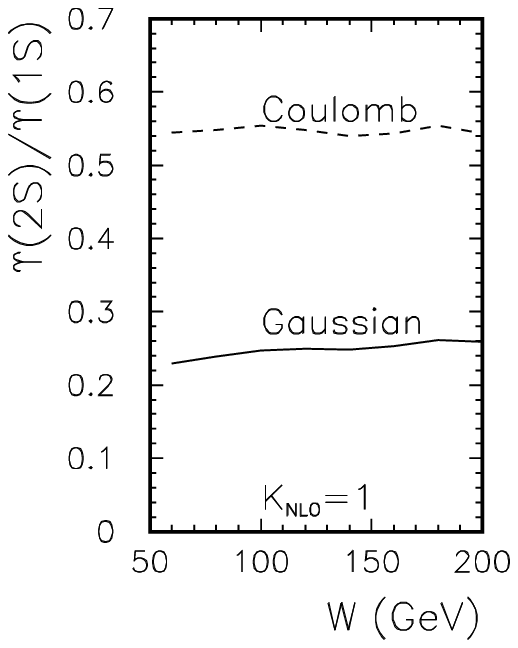}
\includegraphics[width=0.45\textwidth]{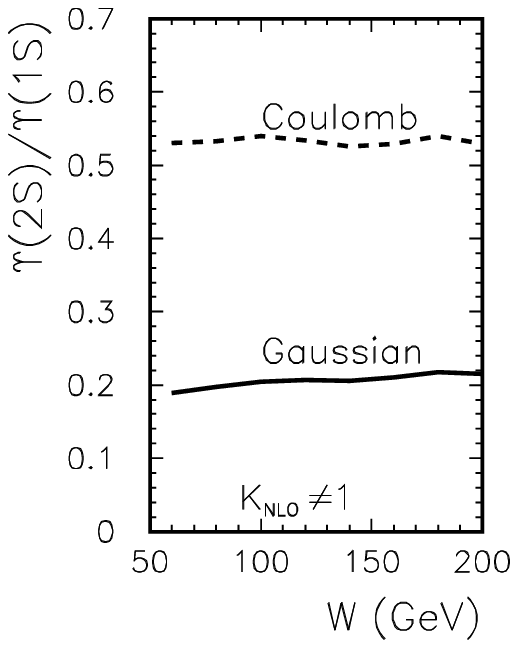}
   \caption{\label{fig:ratio_2S1S}
   \small The $2S/1S$-ratio $\sigma_{tot}(\gamma p \to \Upsilon(2S) p)/ 
\sigma_{tot}(\gamma p \to \Upsilon(1S) p )$ as a function
   of the $\gamma p$ cm--energy. }
\end{figure}

In Fig.\ref{fig:sig_tot} I show the total
cross section for the exclusive $\gamma p \to \Upsilon p$ 
process as a function of the $\gamma p$ cm-energy.
In the left panel I show results for two different 
wave functions discussed in the text: 
Gaussian (solid lines) and Coulomb-like (dashed lines).
Free parameters of the wave function have been
adjusted to reproduce the leptonic decay width in two
ways: (a) using leading order formula (thin lines)
and (b) inlcuding QCD corrections (thick lines).
Including the $K_{NLO}$--factor in the width 
enhances the momentum--space integral over the wave
function (the WF at the spatial origin), and hence
enhances the prediction for the photoproduction cross
section.
The ratio of the cross section for the first radial 
excitation $\Upsilon(2S)$ to the cross section for 
the ground state $\Upsilon(1S)$ is shown in 
Fig.\ref{fig:ratio_2S1S}.
The principal reason behind the suppression of the
$2S$ state is the well--known node effect -- a
cancellation of strength in the $2S$ case due to the 
change of sign of the radial wave function.
It is not surprising, that the numerical
value of the $2S/1S$--ratio is strongly sensitive to the
shape of the radial light--cone wave function.

In our calculations we assumed an equality of the slopes 
for $\Upsilon(1S)$ and $\Upsilon(2S)$ production. 
This appears to be justified, given 
the large spread of predictions from different wave functions.
We finally note, that the ratio depends very little on the 
choice of the $K_{NLO}$ factor (compare left and right panel).

\subsection{Exclusive production of the $\pi^+ \pi^-$ 
pairs\\
in proton-proton collisions}

Up to now I have studied only exclusive production of a 
single mesons. Also the channels with meson pairs
seem interesting. In particular, the channel with
two charged pions which seem feasible experimentally.

\begin{figure}
\begin{center}
  \includegraphics[width=4.5cm]{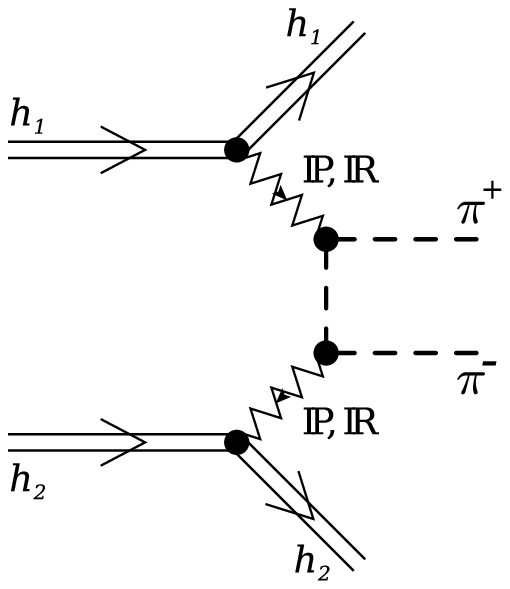}
  \includegraphics[width=4.5cm]{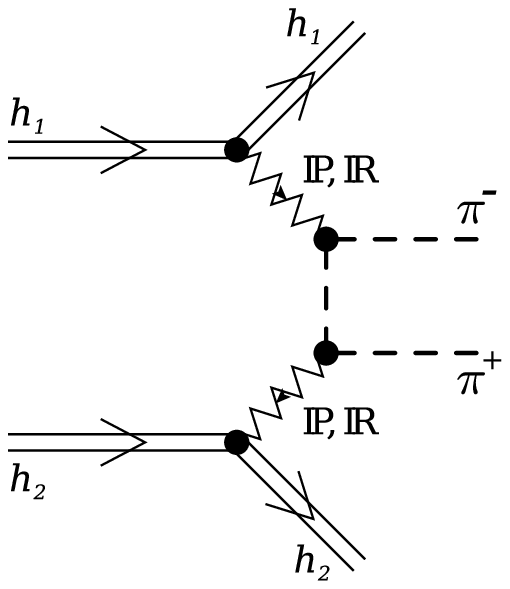}
\end{center}
  \caption{A sketch of the dominant mechanisms of
exclusive production of the $\pi^+ \pi^-$ pairs
at high energies.
\label{fig:2pi_mechanisms}
}
\end{figure}

The underlying mechanism was proposed long ago in 
Ref.\cite{PH76}. The general situation is sketched 
in Fig.\ref{fig:2pi_mechanisms}.
The corresponding amplitude for the
$p p \to p p \pi^+ \pi^-$ process (with four-momenta 
$p_a + p_b \to p_1 + p_2 + p_3 + p_4$) can be written
as
\begin{eqnarray}
{\cal M}^{p p \to p p \pi \pi} &=&
M_{13}(t_1,s_{13}) \; F(t_a) \; 
\frac{1}{t_a - m_{\pi}^2} \;
F(t_a) \; M_{24}(t_2,s_{24}) \nonumber \\
&+&
M_{14}(t_1,s_{14}) \; F(t_b) \;
\frac{1}{t_b - m_{\pi}^2} \;
F(t_b) \; M_{23}(t_2,s_{13})
 \;,
\label{Regge_amplitude}
\end{eqnarray}
where $M_{ik}$ denotes "interaction" between nucleon $i$=1 (forward nucleon) or
$i$=2 (backward nucleon)
and one of the two pions $k=\pi^+$ (3), $\pi^-$ (4). 
In the Regge phenomenology they can be written as:
\begin{eqnarray}
M_{13} &=& s_{13} 
   \left(  C_R^{13} \left(\frac{s_{13}}{s_0}\right)^{\alpha_R-1} \; e^{\frac{B_{\pi N}}{2} \; t_1}
        +  C_P \left(\frac{s_{13}}{s_0}\right)^{\alpha_P-1} \; e^{\frac{B_{\pi N}}{2} \; t_1}
   \right) \; ,
\nonumber \\
M_{14} &=& s_{14}
   \left(  C_R^{14} \left(\frac{s_{14}}{s_0}\right)^{\alpha_R-1} \; e^{\frac{B_{\pi N}}{2} \; t_1}
        +  C_P \left(\frac{s_{14}}{s_0}\right)^{\alpha_P-1} \; e^{\frac{B_{\pi N}}{2} \; t_1}
   \right) \; ,
\nonumber \\
M_{24} &=& s_{23} 
   \left(  C_R^{24} \left(\frac{s_{24}}{s_0}\right)^{\alpha_R-1} \; e^{\frac{B_{\pi N}}{2} \; t_2}
        +  C_P \left(\frac{s_{24}}{s_0}\right)^{\alpha_P-1} \; e^{\frac{B_{\pi N}}{2} \; t_2}
   \right) \; ,
\nonumber \\
M_{23} &=& s_{24}
   \left(  C_R^{23} \left(\frac{s_{23}}{s_0}\right)^{\alpha_R-1} \; e^{\frac{B_{\pi N}}{2} \; t_2}
        +  C_P \left(\frac{s_{23}}{s_0}\right)^{\alpha_P-1} \; e^{\frac{B_{\pi N}}{2} \; t_2}
   \right) \; .
\label{Regge_propagators}
\end{eqnarray}
Above $s_{ik} = W_{ik}^2$, where $W_{ik}$ is the 
center-of-mass energy in the (i,k) subsystem.
The first terms describe the subleading reggeon exchanges 
while the second terms describe exchange of the 
leading (pomeron) trajectory.
The strength parameters of the $\pi N$ interaction
are taken from Ref.\cite{DL92}.
More details of the calculation will be presented 
elsewhere \cite{LS09}.
The $2 \to 4$ amplitude (\ref{Regge_amplitude}) is used
to calculate the corresponding cross section including
limitations of the four-body phase-space.

\begin{figure}[!h]    %
\begin{center}
\includegraphics[width=6cm]{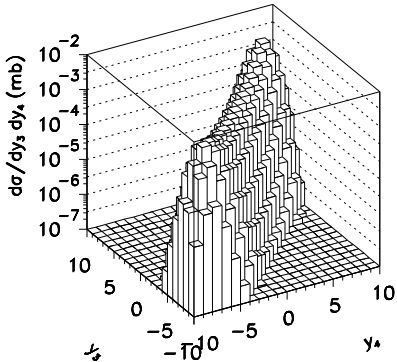}
\end{center}
   \caption{\label{fig:map_y3y4.eps}
   \small 
Rapidity distribution of $\pi^+$ versus $\pi^-$
for W = 14 TeV.
}
\end{figure}

Here I wish to show only one example of the 
two-dimensional distribution in rapidity of positively
charged pion and rapidity of negatively charged pion
at the LHC energy of $W =$ 14 TeV.
The distribution observed differs considerably from
the distribution of the phase space factor.
One can see a two-dimensional shape of the ridge form
elongated along the line $y_3 = y_4$. The minimum of 
the cross section on the top of the ridge occurs when 
$y_3 = y_4 = 0$ and two maxima close to the phase space
ends. The minimum occurs in the part of the phase space
where the pomeron-pomeron contribution dominates, i.e. when both $W_{ik}$ are comparable and large.
The maxima are related to the dominance of
the pomeron-reggeon and reggeon-pomeron mechanisms, i.e.
where one of $W_{ik}$ is small and the second one is large.
The reggeon-reggeon contribution is completely negligible
which is due to the fact that both $W_{ik}$ cannot be
small simultaneously.
We hope that the ALICE collaboration at the LHC will be
able to measure such distributions.

\subsection{Exclusive $A A \to A A \rho^0 \rho^0$  \\
in ultrarelativistic collisions}

Exclusive production of elementary particles (lepton pairs,
Higgs, etc.) or mesons (vector mesons, pair of pseudoscalar
mesons, etc.) in ultrarelativistic collisions is 
an interesting 
and quickly growing field \cite{BGMS75,BHTSK02,Hencken} of 
theoretical investigation.
On experimental side the situation is slightly different.
So far only single-$\rho^0$ exclusive cross section
$A A \to A A \rho^0$ was measured \cite{STAR_rho0}. 
Here the dominant mechanism is a photon-pomeron 
(pomeron-photon) fusion.

\begin{figure}[!h]   
\begin{center}
\includegraphics[width=4cm]{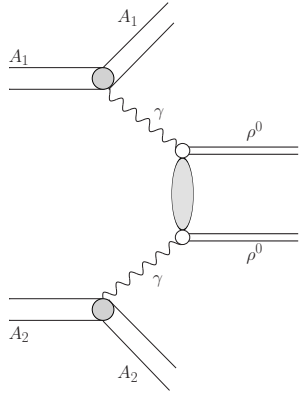}
\end{center}
   \caption{\label{fig:diagram}
   \small 
The reaction discussed in this paper.
}
\end{figure}

Let us consider the process $A A \to A A \rho^0 \rho^0$
depicted in Fig.\ref{fig:diagram}.
The cross section takes the familiar form of a convolution
of equivalent photon fluxes and 
$\gamma \gamma$--cross sections:
\begin{eqnarray}
{d\sigma(AA \to \rho^0\rho^0 AA; s_{AA}) \over d^2 \bb} = 
dn_{\gamma \gamma}(x_1,x_2,\bb) \, {\hat \sigma}(\gamma \gamma \to \rho^0 \rho^0; 
x_1 x_2 s_{AA}) + \dots
\label{nuclear_naive}
\end{eqnarray}
The effective photon flux is expressed through the 
electric field strengths of the ions \cite{KSS09}.

Often flux factors of equivalent, almost on-shell, 
photons are calculated as for point-like particles 
with rescaled charge $e \to Z e$,
and the total cross section is calulated using
a simple parton--model type formula:
\begin{equation}
\sigma \left( AA \to A (\rho^0 \rho^0) A \right) =
\int d \omega_1 d \omega_2 
\frac{n(\omega_1)}{\omega_1}
\frac{n(\omega_2)}{\omega_2}
\hat{\sigma} \left( \gamma \gamma \to \rho^0 \rho^0 \right) \; .
\label{EPA_formula}
\end{equation}
The formulae (\ref{EPA_formula})
clearly does not take 
into account absorption effects when initial nuclei 
undergo nuclear breakup. This can be easily done in 
the impact parameter space where the geometry of the
collision is more explicit.
Then rather two-dimensional flux factors \cite{Jackson} 
must be used.

The simple EPA formula can be generalized to
\begin{eqnarray}
&&\sigma \left( AA \to A (\rho^0 \rho^0) A \right) =
\int d^2 b_1 d \omega_1 d^2 b_2 d \omega_2 
N(\omega_1, b_1)
N(\omega_2, b_2) \; 
\nonumber \\
&& \theta 
\left(|\vec{b}_1 - \vec{b}_2| - R_{12} \right) \;
{\hat \sigma} 
\left( \gamma \gamma \to \rho^0 \rho^0 \right) \; .
\label{bspace_EPA_formula}
\end{eqnarray}
Here an extra $\theta$ function was introduced
which excludes those cases when nucler collisions,
leading to nuclear breakup, take place 
($R_{12} = R_1 + R_2$).
The two-dimensional fluxes in (\ref{bspace_EPA_formula}) 
can be calculated in terms of the charge form factor
of nucleus \cite{BF91} as:
\begin{equation}
N(w,b) = \frac{Z^2 \alpha}{\pi^2} \Phi(x,b) \; ,
\label{2dim_flux}
\end{equation}
where the auxiliary function $\Phi$ reads: 
\begin{equation}
\Phi(x,b) = \Big|
\int_{0}^{\infty} du \; u^2 J_1(u) \;
\frac{F(-(x^2+u^2)/b^2)}{x^2 + u^2} 
\Big|^2  \; .
\label{auxiliary_Phi}
\end{equation}
The second ingredient of our approach is 
$\gamma \gamma \to \rho^0 \rho^0$ cross section.
Here the situation is not well established.
The cross section for this process was measured
up to $W_{\gamma \gamma}$ = 4 GeV \cite{MPW}.
At low energy one observes a huge increase of the cross section.

\begin{figure}[!h]   
\begin{center}
\includegraphics[width=0.4\textwidth]{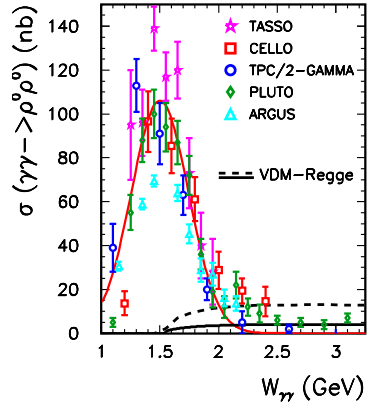}
\end{center}
   \caption{\label{fig:fit_gamgam_rhorho}
   \small
The elementary cross section for 
the $\gamma \gamma \to \rho^0 \rho^0$ reaction.
In the left panel we display the collection of the $e^+ e^-$ 
experimental data \cite{MPW} and our fit.
In the right panel we show our predictions based on the VDM-Regge
model decribed in the text. For comparison we show also result when
the form factor correcting for off-shell effect is ignored
(see \cite{KSS09}).
}
\end{figure}

In Fig.\ref{fig:fit_gamgam_rhorho} we have collected 
the world data (see \cite{MPW} 
and references therein).
We use rather directly experimental data in order 
to evaluate the cross section in nucleus-nucleus 
collisions. In Fig.\ref{fig:fit_gamgam_rhorho} we 
show our fit to the world data
.

\begin{figure}[!h]  
\begin{center}
\includegraphics[width=0.4\textwidth]{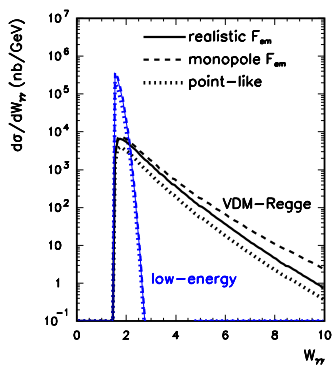}
\end{center}
   \caption{\label{fig:dsig_dW} 
  \small 
The $Au + Au \to Au + Au + \rho^0 \rho^0$ cross section
as a function of $W_{\gamma \gamma}$ = $M_{\rho \rho}$
for the RHIC energy $\sqrt{s}_{NN}$ = 200 GeV.
The low- and high-energy components are shown separately.
}
\end{figure}

The cross section above $W$ = 4 GeV was never measured
in the past. It is well known that the cross section 
for $\gamma \gamma \to$ hadrons can be well described
in the VDM-Regge type model.
We use a similar approach for the
final state channel $\rho^0 \rho^0$.
In Fig.\ref{fig:fit_gamgam_rhorho} we present
the corresponding $t$-integrated cross section together
with existing experimental data taken from 
\cite{MPW}.
The vanishing of the VDM-Regge cross section at 
$W_{\gamma \gamma} = 2 m_{\rho}$ is due to 
$t_{min}$, $t_{max}$ limitations.
It is obvious from Fig.\ref{fig:fit_gamgam_rhorho} that the VDM-Regge 
model cannot explain the huge close-to-threshold enhancement. 
In Fig.\ref{fig:dsig_dW} we show distribution
of the cross section for the nucleus-nucleus scattering
in photon-photon center-of-mass energy for
both low-energy component and high-energy 
VDM-Regge component. Below $W$ = 2 GeV the low-energy
component dominates. The situation reverses above
$W$ = 2 GeV. One can study the high energy component
imposing extra cut on $M_{\rho \rho}$.
However, the cross section drops quickly with increasing
invariant mass of two-$\rho$ mesons.

\begin{figure}[!h]   
\begin{center}
\includegraphics[width=0.4\textwidth]{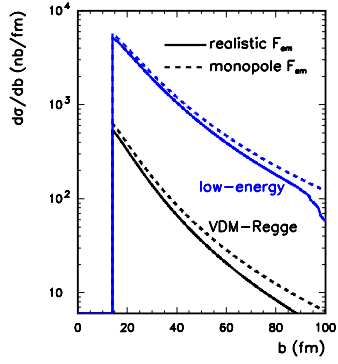}
\end{center}
   \caption{\label{fig:dsig_dbm}
   \small 
The $Au + Au \to Au + Au + \rho^0 \rho^0$ cross section
as a function of the impact parameter $b$ for 
$\sqrt{s}_{NN}$ = 200 GeV. The meaning of the curves is
the same as in Fig.\ref{fig:dsig_dW}.
The cut off for $R_{12} \approx$ 14 fm is clearly visible.
}
\end{figure}

For illustration in Fig.\ref{fig:dsig_dbm} we show 
the model distribution in impact parameter 
$b = |\vec{b}_1 - \vec{b}_2|$.
We show distributions for the low- and high-energy
components separately. I also show distributions
for point-like charge, monopole form factor
and realistic charge density (see \cite{KSS09}).
One can see slightly different results for different 
approaches how to calculate flux factors of equivalent
photons.

\begin{figure}[!h]  
\includegraphics[width=0.4\textwidth]{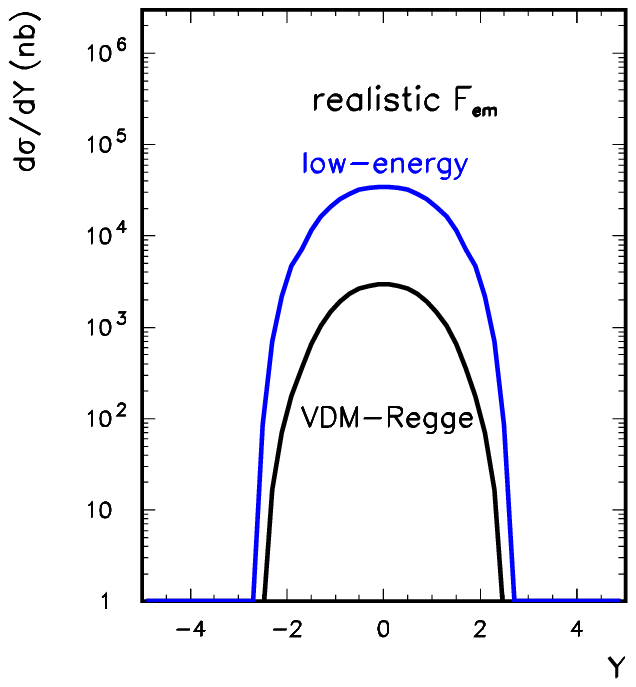}
\includegraphics[width=0.4\textwidth]{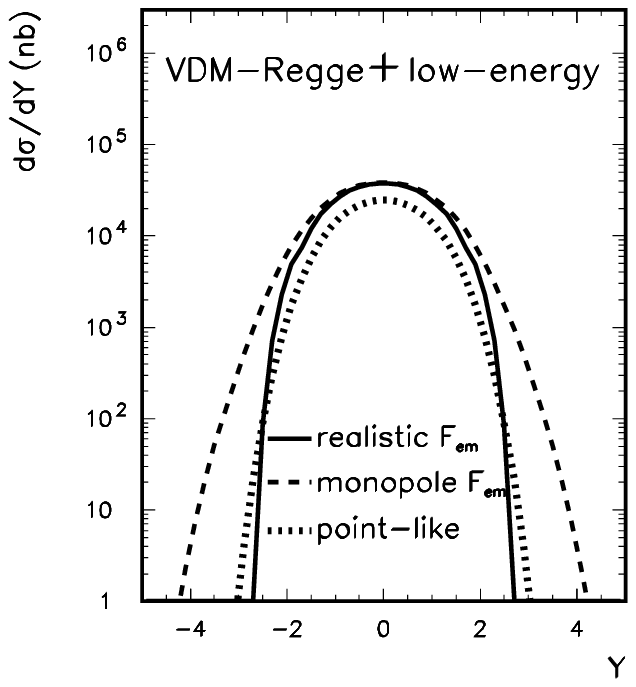}
   \caption{\label{fig:dsig_dY}
   \small 
The $Au + Au \to Au + Au + \rho^0 \rho^0$ cross section
as a function of the rapidity of the $\rho^0 \rho^0$ pair 
$Y$ for $\sqrt{s}_{NN}$ = 200 GeV.
The meaning of the curves is
the same as in Fig.\ref{fig:dsig_dW}
}
\end{figure}

Finally in Fig.\ref{fig:dsig_dY} I show distribution
in rapidity of the $\rho^0 \rho^0$ pair.
Compared to the point-like case, the distribution 
obtained with realistic charge density is concentrated
at midrapidities, and configurations when both $\rho^0$'s
are in very forward or both $\rho^0$'s are in very backward
directions are strongly damped compared to the case
with point-like nucleus charges. One can also see 
a difference between results obtained with 
an approximate monopole form factor 
and with the exact one calculated from realistic charge 
density.

\section{Conclusions}

We have derived, for the first time in the literature,
QCD amplitude for exclusive elastic double diffractive
production of axial-vector $\chi_c(1^+)$ meson.
According to the Landau-Yang theorem the amplitude
vanishes for the fusion of on-shell gluons.
We have generalized the formalism proposed recently for 
diffractive production of the Higgs boson and derived
corresponding $g^* g^* \to \chi_c(1^+)$ vertex function.
The predicted total cross section, obtained from the 
bare amplitude is of the order of a fraction of
nb, depending on the model of UGDF.
This is about two orders of magnitude less than a similar
cross section for $\chi_c(0^+)$.

However, because the branching fraction $BR(\chi_c(1^+) \to
J/\psi + \gamma) \gg BR(\chi_c(0^+) \to J/\psi + \gamma)$,
one may expect a different situation in
the $J/\psi + \gamma$ channel.
This is similar as for inclusive production of
P-wave quarkonia, where the signal
(in the $J/\psi + \gamma$ channel) of $\chi_c(1^+)$
is of similar size as that for $\chi_c(0^+)$.

We have calculated the forward amplitude for 
$\gamma p \to \Upsilon p$ reaction within the formalism 
of $k_t$-factorization. In this approach
the energy dependence of the process is encoded in 
the $x$-dependence of unintegrated gluon distributions. 
The latter object is constrained by data 
on inclusive deep inelastic scattering.
The $t$-dependence for the $\gamma p \to \Upsilon p$ 
process involves a free parameter and is parametrized. 
We have used different Ans\"atze for the $b \bar b$ 
wave functions.
The results for $\Upsilon(1S)$ production depend only 
slightly on the model of the wave function, while 
the $2S/1S$ ratio shows a substantial sensitivity. 
We compared our results for the total cross section 
with a recent data from HERA. 
Our results are somewhat lower than data,
although the overall discrepancy is not worrysome, 
given the large uncertainties due to the rather poor 
experimental resolution in the meson mass. 
The amplitudes for the $\gamma p \to \Upsilon p$ process 
are used next to calculate the amplitude for the 
$p \bar p \to p \bar p \Upsilon$ reaction assuming 
the photon-Pomeron (Pomeron-photon) underlying dynamics.
We have calculated several differential distributions 
including soft absorption effects not included so far 
in the literature.
Our predictions are relevant for current experiments 
at the Tevatron.

For the first time in the literature we have estimated the cross 
section for exclusive $f_0(1500)$ meson (glueball candidate) production 
not far from the threshold.
We have included both gluon induced diffractive
mechanism and the pion-pion exchange contributions. 

The first component was obtained by extrapolating down the cross section
in the Khoze-Martin-Ryskin approach with unintegrated gluon distributions 
from the literature as well as using two-gluon impact factor approach. 
A rather large uncertainties are associated with 
the diffractive component.
At present only upper limit can be obtained for the diffractive
component as the $f_0(1500) \to g g$ decay coupling constant remains 
unknown. The coupling constant could be
extracted only in high-energy exclusive production of $f_0(1500)$
where other mechanisms are negligible.

The calculation of the meson-exchange contribution requires
introducing extra vertex form factors.
At largest PANDA energies they are relatively well known and
the pion-pion fusion can be reliably calculated.
We predict the dominance of the pion-pion contribution
close to the threshold. Our calculation shows that the diffractive
component is by more than order of magnitude
smaller than the pion-pion fusion component in the energy 
region of the future PANDA experiments.
The diffractive component may dominate over the pion-pion component
only for center-of-mass energies $W >$ 15 GeV.
Clearly an experimental program is required to disentagle 
the reaction mechanism.

We have made first estimate of the exclusive production
of pairs of charged pions at high-energy. 
Different combinations of pomeron-reggeon fusion
were included.
Rather large cross sections are predicted at RHIC,
Tevatron and LHC. 
Here I have shown only rapidity distributions of pions
at LHC.
The production of the two pions is strongly correlated
in the ($y(\pi^+), y(\pi^-)$) space. 
Even at relatively high energies
the inclusion of reggeon exchanges is crucial as
amplitudes with different combination of exchanges
interfere or/and $\pi N$ subsystem energies can be 
relatively small $W_{\pi N} <$ 10 GeV.
At high-energies we find a preference for the same 
hemisphere (same-sign rapidity) emission of 
$\pi^+$ and $\pi^-$. For example at LHC energies 
the same hemisphere emission constitutes about 90 \% 
of all cases.

We have calculated, for the first time, realistic 
cross sections for exclusive $\rho^0 \rho^0$ production 
in ultrarelativistic heavy-ion collisions at RHIC. 
We have used realistic charge densities to calculate
the nuclear charge form factors.
The absorption effects have been included.

It was shown that calculating both flux factors and
$\gamma \gamma \to \rho^0 \rho^0$
realistically is necessary to make reliable etimates 
of the nucleus-nucleus exclusive production of the 
$\rho^0 \rho^0$ pairs.
Large cross sections, of the order of fraction of milibarn,
have been found. The bulk of the cross section is, however,
concentrated in low photon-photon energies 
(low $\rho^0 \rho^0$ invariant masses)

The $\rho^0$ mesons, decaying into $\pi^+ \pi^-$, can 
be measured e.g. by the STAR detector at RHIC and 
the ALICE detector at LHC.

\vspace{1cm}

{\bf Acknowledgments}
The collaboration with Wolfgang Sch\"afer, Roman Pasechnik,
Oleg Teryaev, Anna Cisek, Mariola K{\l}usek and 
Piotr Lebiedowicz on the topics presented here 
is acknowledged. This work was partially supported by the 
Polish Ministry of Science and Higher Education
under grants no. N N202 249235 and N N202 078735.

\end{document}